\documentclass[11pt]{article}
\usepackage{amsfonts}
\usepackage[letterpaper]{geometry}
\usepackage{amssymb}
\usepackage{amsmath}
\usepackage{array}
\usepackage{slashed}
\usepackage{hyperref}
\usepackage{color}
\usepackage{graphicx}
\usepackage{float}
\usepackage{lmodern}
\usepackage[utf8]{inputenc}
\usepackage{cite}

\pdfoutput=1

\numberwithin{equation}{section}
\begin{document}

%TCIMACRO{\TeXButton{B titlepage}{\begin{titlepage}}}%
%BeginExpansion
\begin{titlepage}%
%EndExpansion

\vspace*{40pt}

\begin{center}
{\LARGE \ \textbf{Higher-spin initial data in twistor space with complex
stargenvalues}\ }

\bigskip

\bigskip

{\Large Yihao Yin} \ \let\thefootnote\relax\footnote{%
yinyihao@nuaa.edu.cn , \ yinyihao@gmail.com}

\bigskip

\bigskip

\textit{College of Science, Nanjing University of Aeronautics and
Astronautics, }

\textit{Jiangjun Avenue 29, Nanjing, China }

\textit{\bigskip }

\textit{\& }

\textit{\bigskip }

\textit{Departamento de Ciencias F\'{\i}sicas, Universidad Andr\'{e}s Bello, 
}

\textit{Rep\'{u}blica 220, Santiago de Chile}

\bigskip

\bigskip

\bigskip

\bigskip

%TCIMACRO{\TeXButton{B abstract}{\begin{abstract}} }%
%BeginExpansion
\begin{abstract}
%EndExpansion
This paper is a supplement to and extension of arXiv:1903.01399. In the
internal twistor space of the 4D Vasiliev's higher-spin gravity, we study
the star-product eigenfunctions of number operators with generic complex
eigenvalues. In particular, we focus on a set of eigenfunctions represented
by formulas with generalized Laguerre functions. This set of eigenfunctions
can be written as linear combinations of two subsets of eigenfunctions, one
of which is closed under the star-multiplication with the creation operator
to a generic complex power -- and the other similarly with the annihilation
operator. The two subsets intersect when the left and the right eigenvalues
differ by an integer. We further investigate how star-multiplications with
both the creation and annihilation operators together may change such
eigenfunctions and briefly discuss some problems that we are facing in order
to use these eigenfunctions as the initial data to construct solutions to
Vasiliev's equations. 
%TCIMACRO{\TeXButton{E abstract}{\end{abstract}}}%
%BeginExpansion
\end{abstract}%
%EndExpansion
\end{center}

%TCIMACRO{\TeXButton{E titlepage}{\end{titlepage}}}%
%BeginExpansion
\end{titlepage}%
%EndExpansion

\setcounter{footnote}{0}

\section{Introduction}

Vasiliev's equations of higher-spin gravity \cite%
{Vasiliev:1990en,Vasiliev:1999ba} are known as an interacting theory with
infinitely many higher-spin fields, which can be seen as a non-linear
extension of (Fang)-Fronsdal equations of free higher-spin fields \cite%
{Fronsdal:1978rb,Fang:1978wz}. General relativity, which in some sense can
be seen as a non-linear extension of the spin-2 Fronsdal equation, has much
richer physical content than merely spin-2 particles. Likewise, Vasiliev's
equations are not just about higher-spin particles, and in order to
comprehend various aspects of their physical implications, efforts have been
made to find and interpret their solutions \cite%
{Sezgin:2005pv,Iazeolla:2007wt,Didenko:2009td,Iazeolla:2017vng,Iazeolla:2017dxc,Iazeolla:2011cb,Iazeolla:2012nf,Gubser:2014ysa,Bourdier:2014lya,Sundell:2016mxc,Aros:2017ror,Aros:2019pgj}%
.\footnote{%
In this paper, we only focus on the 4D spacetime, though efforts have also
been made to solve the 3D Prokushkin-Vasiliev theory \cite%
{Prokushkin:1998bq,Didenko:2006zd,Iazeolla:2015tca}.}

A frequently used method to solve Vasiliev's equations, which was proposed
in \cite{Vasiliev:1990bu,Sezgin:2005pv}, is that we can first construct
solutions in the absence of ordinary spacetime (or more often referred to as
\textquotedblleft initial data\textquotedblright\ of solutions), and then
turn on spacetime by doing gauge transformations (see \cite%
{Iazeolla:2017dxc,DeFilippi:2019jqq} for recent review and development).

To illustrate this method, let us look at the Vasiliev's equations in 4D
spacetime for all integers spins, written in their component equations: 
\begin{subequations}
\label{Vaseq}
\begin{eqnarray}
\partial _{\lbrack \mu }U_{\nu ]}+U_{[\mu }\star U_{\nu ]} &=&0\text{ ,} \\
\partial _{\mu }\Phi +U_{\mu }\star \Phi -\Phi \star \pi \left( U_{\mu
}\right) &=&0\text{ ,} \\
\partial _{\mu }V_{\alpha }-\partial _{\alpha }U_{\mu }+\left[ U_{\mu
},V_{\alpha }\right] _{\star } &=&0\text{\ \ \ \ and \ \ h.c. ,} \\
\partial _{\lbrack \alpha }V_{\beta ]}+V_{[\alpha }\star V_{\beta ]}+\tfrac{i%
}{4}\varepsilon _{\alpha \beta }\Phi \star \kappa &=&0\text{\ \ \ \ and \ \
h.c. ,} \\
\partial _{\alpha }\Phi +V_{\alpha }\star \Phi -\Phi \star \bar{\pi}\left(
V_{\alpha }\right) &=&0\text{\ \ \ \ and \ \ h.c. ,} \\
\partial _{\alpha }\bar{V}_{\dot{\alpha}}-\partial _{\dot{\alpha}}V_{\alpha
}+\left[ V_{\alpha },\bar{V}_{\dot{\alpha}}\right] _{\star } &=&0\text{ ,}
\end{eqnarray}%
where all the fields can depend on three sets of coordinates: $x^{\mu }$ for
the ordinary 4D spacetime, $Y^{\underline{\alpha }}$ and $Z^{\underline{%
\alpha }}$ for the two 4-dimensional internal and external symplectic
manifolds (often referred to as \textquotedblleft twistor
spaces\textquotedblright ), which can be decomposed as $Y^{\underline{\alpha 
}}=(y^{\alpha },\bar{y}^{\dot{\alpha}})$ and $Z^{\underline{\alpha }%
}=(z^{\alpha },\bar{z}^{\dot{\alpha}})$ with $(y^{\alpha })^{\dag }=\bar{y}^{%
\dot{\alpha}}$ and $(z^{\alpha })^{\dag }=-\bar{z}^{\dot{\alpha}}$. The
underlined indices are Sp(4,$\mathbb{R}$) or USp(2,2) indices for the AdS or
dS background, and $\{\alpha ,\beta ,\cdots \}$ are SL(2,$\mathbb{C}$)
indices raised or lowered by the Levi-Civita symbols $\varepsilon ^{\alpha
\beta }$, $\varepsilon _{\alpha \beta }$, $\varepsilon ^{\dot{\alpha}\dot{%
\beta}}$ or $\varepsilon _{\dot{\alpha}\dot{\beta}}$. The SL(2,$\mathbb{C}$)
indices that appear in (\ref{Vaseq}) refer to those of $Z$-coordinates,
while the $Y$-coordinates are building blocks of symmetry algebra
generators, which are implicit in (\ref{Vaseq}) and whose indices are
contracted with the ones of field components. In the equations, $\Phi $ is a
zero-form, $U_{\mu }$ and $V_{\alpha }$ (with its hermitian conjugate $\bar{V%
}_{\dot{\alpha}}$) are the spacetime and twistor-space components of a
one-form field, functioning as gauge fields. The star-product between the
fields can be formally defined as 
\end{subequations}
\begin{eqnarray}
&&f_{1}\left( y,\bar{y},z,\bar{z}\right) \star f_{2}\left( y,\bar{y},z,\bar{z%
}\right)  \notag \\
&&\ =\ \int d^{2}ud^{2}\bar{u}d^{2}vd^{2}\bar{v}\ (2\pi )^{-4}\ e^{i\left(
v^{\alpha }u_{\alpha }+\bar{v}^{\dot{\alpha}}\bar{u}_{\dot{\alpha}}\right)
}\   \notag \\
&&\ \ \ \ \ \ \ \ f_{1}\left( y+u,\bar{y}+\bar{u};z+u,\bar{z}+\bar{u}\right)
f_{2}\left( y+v,\bar{y}+\bar{v};z-v,\bar{z}-\bar{v}\right) \text{ .}
\end{eqnarray}%
Furthermore, $\kappa $ is the inner Klein operator satisfying 
\begin{equation}
\kappa =\kappa _{y}\star \kappa _{z}\text{ , \ }\kappa _{y}=2\pi \delta
^{2}(y)\text{ , \ }\kappa _{z}=2\pi \delta ^{2}(z) \text{ ,}
\end{equation}%
(idem.\ $\bar{\kappa}$), where $\delta ^{2}$ is the two dimensional Dirac
delta function, and $\pi $ and $\bar{\pi}$ are operations that flip the
signs of $Y$ and $Z$ coordinates:%
\begin{eqnarray}
\pi \left( x^{\mu };y^{\alpha },\bar{y}^{\dot{\alpha}};z^{\alpha },\bar{z}^{%
\dot{\alpha}}\right) &=&\left( x^{\mu };-y^{\alpha },\bar{y}^{\dot{\alpha}%
};-z^{\alpha },\bar{z}^{\dot{\alpha}}\right) \text{ ,}  \notag \\
\bar{\pi}\left( x^{\mu };y^{\alpha },\bar{y}^{\dot{\alpha}};z^{\alpha },\bar{%
z}^{\dot{\alpha}}\right) &=&\left( x^{\mu };y^{\alpha },-\bar{y}^{\dot{\alpha%
}};z^{\alpha },-\bar{z}^{\dot{\alpha}}\right) \text{ .}
\end{eqnarray}%
For the bosonic model, the fields should satisfy $\pi \bar{\pi}\left( \Phi
,U_{\mu },V_{\alpha }\right) =\left( \Phi ,U_{\mu },-V_{\alpha }\right) $,
so that non-integer-spin degrees of freedom are projected out, and depending
on the background being AdS or dS, different reality conditions should be
imposed on these fields (see \cite{Iazeolla:2007wt,Aros:2017ror} for
details, which we will skip here).

The simplest solutions to (\ref{Vaseq}) are vacuum solutions. As can be
easily seen, if we set all fields to zero except the spacetime gauge field $%
U_{\mu }$, then $U_{\mu }$ becomes a pure-gauge independent of $Z$%
-coordinates, and thus 
\begin{equation}
U_{\mu }=L^{-1}\star \partial _{\mu }L\text{ ,}  \label{ULL}
\end{equation}%
with an arbitrary gauge function $L\left( x,Y\right) $ is a solution.
Whether such a solution is AdS or dS or something else depends on the
expression of $L$.

To obtain solutions other than the vacua, we can do things in the opposite
way: if we trivialize spacetime i.e.\ set $U_{\mu }$ to zero and let $\Phi $
and $V$ be independent of $x^{\mu }$, then the first three equations in (\ref%
{Vaseq}) are directly solved and thus we can focus on solving the last three
equations without involving spacetime (similar to the method of separating
variables for solving differential equations). In other words, (\ref{Vaseq})
can be solved by using the following ansatz: 
\begin{eqnarray}
U_{\mu } &=&G^{-1}\star \partial _{\mu }G  \notag \\
\Phi &=&G^{-1}\star \Phi ^{\prime }\star \pi (G)  \notag \\
V_{\alpha } &=&G^{-1}\star V_{\alpha }^{\prime }\star G+G^{-1}\star \partial
_{\alpha }G\text{ ,}  \label{GAnsatz}
\end{eqnarray}%
where the primed fields are independent of spacetime. After we obtain the
solution for the primed fields, we can turn on spacetime by properly doing a
gauge transformation\footnote{%
Due to the $\pi $,$\bar{\pi}$-automorphisms of the symmetry algebra, two
different adjoint representations exist in Vasiliev's equations, namely the
adjoint and the twisted adjoint representations. $\Phi $ lives in the
twisted\ adjoint representation whose gauge transformation is modified with
a $\pi $ as shown in (\ref{GAnsatz}).} with gauge function $G$ that enables
the extraction of Fronsdal fields. We often write $G=L\star H$, where $L$ is
the vacuum gauge function in (\ref{ULL}), so that we can do the gauge
transformation in two steps, because doing the $L$-gauge transformation is
much easier than the full gauge transformation $G$, and with the $L$-gauge
we can already study some spacetime properties of the solution at the
linearized level (see e.g.\ \cite%
{Iazeolla:2017vng,Aros:2017ror,DeFilippi:2019jqq}). Now by substituting (\ref%
{GAnsatz}) into (\ref{Vaseq}), the first three equations of (\ref{Vaseq})
are directly solved and the last three are converted to the same equations
with all the fields primed: 
\begin{subequations}
\label{primedeq}
\begin{eqnarray}
\partial _{\lbrack \alpha }V_{\beta ]}^{\prime }+V_{[\alpha }^{\prime }\star
V_{\beta ]}^{\prime }+\tfrac{i}{4}\varepsilon _{\alpha \beta }\Phi ^{\prime
}\star \kappa &=&0\text{\ \ \ \ and \ \ h.c. ,}  \label{primedeqkappa} \\
\partial _{\alpha }\Phi ^{\prime }+V_{\alpha }^{\prime }\star \Phi ^{\prime
}-\Phi ^{\prime }\star \bar{\pi}\left( V_{\alpha }^{\prime }\right) &=&0%
\text{\ \ \ \ and \ \ h.c. ,} \\
\partial _{\alpha }\bar{V}_{\dot{\alpha}}^{\prime }-\partial _{\dot{\alpha}%
}V_{\alpha }^{\prime }+\left[ V_{\alpha }^{\prime },\bar{V}_{\dot{\alpha}%
}^{\prime }\right] _{\star } &=&0\text{ .}
\end{eqnarray}%
The last two equations of (\ref{primedeq}) can be directly solved by the
ansatz\footnote{$\Psi $ lives in the adjoint representation and $\Phi
^{\prime }$ lives in the twisted one. The $\star \kappa _{y}$ can be used to
convert between them. \cite{Didenko:2009td} This ansatz was first explicitly
used in \cite{Iazeolla:2017vng}. See also \cite%
{Iazeolla:2017dxc,Aros:2017ror} for details.} 
\end{subequations}
\begin{eqnarray}
\Phi ^{\prime } &=&\Psi \left( y,\bar{y}\right) \star \kappa _{y}\text{ ,} 
\notag \\
V_{\alpha }^{\prime } &=&\sum_{n=1}^{\infty }a_{\alpha }^{(n)}(z)\star \Psi
\left( y,\bar{y}\right) ^{\star n}\text{\ \ with\ \ }\kappa _{z}\star
a_{\alpha }^{(n)}(z)\star \kappa _{z}=-a_{\alpha }^{(n)}(z)\text{ ,}
\label{ansatzholo}
\end{eqnarray}%
and (\ref{primedeqkappa}) after being substituted with this ansatz can also
be solved for $a_{\alpha }^{(n)}$ (see \cite{Iazeolla:2011cb} for details,
which we will not involve in this paper). Here the $V_{\alpha }^{\prime }$
field is expressed as a star-product Taylor series of $\Psi $, where we
label the coefficient of the $n$th-order term by a superscript $(n)$, and it
is a holomorphic function of the $z$-coordinates (thus this particular gauge
for finding solutions is often called the holomorphic gauge). In this gauge
the purely $Y$-dependent expression of $\Psi $ or equivalently $\Phi
^{\prime }$ is like the \textquotedblleft initial data\textquotedblright ,
i.e.\ once $\Psi $ or $\Phi ^{\prime }$ is properly chosen, the solution is
certain.\footnote{%
The solution is certain under the assumption that the expression of the
gauge function $G$ can be fixed by demanding a good physical interpretation
of the solution. See \cite{DeFilippi:2019jqq} for a recent discussion.}

With the above framework settled, finding a solution with desirable
properties is to a large extent the art of constructing the initial data. A
very useful and systematic way of constructing the initial data was given in 
\cite{Iazeolla:2011cb}, where the authors constructed on the $Y$-coordinates
two Fock spaces, whose number operators are the sum and difference of a pair
of Cartan generators of the AdS$_{\text{4}}$ isometry algebra. Linear
combinations of the star-product eigenfunctions (stargenfunctions) of the
number operators are used as the initial data, which after a
spacetime-dependent gauge transformation can turn into large sets of
solutions with spacetime symmetries corresponding to the Cartan generators.
To make this short paper self-contained, let us briefly go through some
technical details of such stargenfunctions.

Let us define the creation and annihilation operators as some constant
linear combinations of the $Y$-coordinates, i.e.\ 
\begin{equation}
a^{+}=A_{\underline{\alpha }}^{+}Y^{\underline{\alpha }}\text{ \ and\ \ }%
a^{-}=A_{\underline{\alpha }}^{-}Y^{\underline{\alpha }}\text{\ \ with \ }%
A^{-\underline{\alpha }}A_{\underline{\alpha }}^{+}=\frac{i}{2}\text{ \ ,}
\label{alinY}
\end{equation}%
so that $\left[ a^{-},a^{+}\right] =1$. The number operator in principle
should be defined as $a^{+}\star a^{-}$, which is equal to $a^{+}a^{-}-\frac{%
1}{2}$, but for convenience we use as in \cite{Iazeolla:2011cb} the shifted
number operator instead (we drop the word \textquotedblleft
shifted\textquotedblright\ from now): 
\begin{equation}
w=a^{+}a^{-}\text{ .}
\end{equation}%
We denote the stargenfunction of the number operator $f_{\lambda
_{L}|\lambda _{R}}\left( a^{+},a^{-}\right) $ with $\lambda _{L}$ and $%
\lambda _{R}$ being the left and the right stargenvalues and assume that the
function depends on the $Y$-coordinates only via the creation and
annihilation operators, and thus 
\begin{subequations}
\label{originaleq}
\begin{eqnarray}
w\star f_{\lambda _{L}|\lambda _{R}}\left( a^{+},a^{-}\right) &=&\lambda
_{L}f_{\lambda _{L}|\lambda _{R}}\left( a^{+},a^{-}\right) \text{ ,}
\label{originaleqL} \\
f_{\lambda _{L}|\lambda _{R}}\left( a^{+},a^{-}\right) \star w &=&\lambda
_{R}f_{\lambda _{L}|\lambda _{R}}\left( a^{+},a^{-}\right) \text{ .}
\label{originaleqR}
\end{eqnarray}%
We can make two copies of the above set of creation, annihilation and number
operators and stargenfunctions (label them with \textquotedblleft
1\textquotedblright\ and \textquotedblleft 2\textquotedblright ), such that $%
w_{2}\pm w_{1}$ are a pair of Cartan generators\footnote{%
The generators are selected among $E$, $J$, $iB$ and $iP$, which are
time-translation, rotation, boost and spatial transvection respectively, and
for non-compact generators $B$ and $P$ the imaginary factor $i$ has to be
multiplied for constructing the Fock spaces.} of the (A)dS isometry algebra,
then a convenient way of constructing the initial data $\Psi $ or $\Phi
^{\prime }$, is to let them be equal to the product between the two copies
of stargenfunctions i.e.\footnote{%
One can prove that the star-product on the r.h.s.\ is equal to the ordinary
product, due to commutativity between the creation, annihilation and number
operators from different copies.} 
\end{subequations}
\begin{equation}
f_{\lambda _{1L},\lambda _{2L}|\lambda _{1R},\lambda _{2R}}\left(
a_{1}^{+},a_{1}^{-},a_{2}^{+},a_{1}^{-}\right) =f_{\lambda _{1L}|\lambda
_{1R}}\left( a_{1}^{+},a_{1}^{-}\right) \star f_{\lambda _{2L}|\lambda
_{2R}}\left( a_{2}^{+},a_{2}^{-}\right) \text{ .}  \label{eigenfunc12}
\end{equation}%
For example, in the case that the energy and angular momentum operators are $%
E=\frac{1}{2}\left( w_{2}+w_{1}\right) $ and $J=\frac{1}{2}\left(
w_{2}-w_{1}\right) $ and the stargenfunctions live in the adjoint
representation, if we use diagonal stargenfunctions i.e. set\ $\lambda
_{1L}=\lambda _{1R}=\lambda _{1}$ and $\lambda _{2L}=\lambda _{2R}=\lambda
_{2}$, which leads to vanishing commutators $\left[ E,f_{\lambda
_{1},\lambda _{2}|\lambda _{1},\lambda _{2}}\right] _{\star }$ and $\left[
J,f_{\lambda _{1},\lambda _{2}|\lambda _{1},\lambda _{2}}\right] _{\star }$,
then we expect the solution after switching on spacetime by a gauge
transformation should exhibit the symmetries of the time-translation and the
rotation along a certain spatial axis.

The paper \cite{Iazeolla:2011cb} only allowed the stargenvalues $\lambda $'s
to be half-integers and mainly focused on diagonal cases, which seemed to be
too much constraint later when we studied fields on BTZ-like backgrounds. It
is well-known that the 3D BTZ black hole \cite{Banados:1992wn} can be
obtained from AdS$_{\text{3}}$ by compactifying the non-compact spatial
direction corresponding to a spatially transvectional isometry \cite%
{Banados:1992gq}. Field fluctuations in the context of Vasiliev's equations
on such a 3D background were studied in \cite{Didenko:2006zd}, and in \cite%
{Aros:2019pgj} we tried to study them in 4D instead with a BTZ-like
background obtained from AdS$_{\text{4}}$ by the same kind of
compactification \cite{Aminneborg:1996iz,Holst:1997tm,Banados:1998dc}. Such
compactification leads to a periodicity condition on the fields, which leads
to the requirement in the initial data that the stargenfunctions of the
spatial transvection generator $iP$ should in general be non-diagonal with
complex stargenvalues, and furthermore, the excitation of the (angular)
momentum along the compactified direction corresponds to adding quantized
imaginary numbers to the stargenvalues in the initial data.

This motivates us to systematically study the solutions to (\ref{originaleq}%
) in a more general sense with arbitrary complex left and right
stargenvalues and to investigate how far we can go to organize these
stargenfunctions into Fock spaces. In Section \ref{Sec solveeigen}, we
present solutions to (\ref{originaleq}) in terms of special functions, which
are converted into integral representations in Section \ref{Sec integral}
for convenience of doing star-products. Then in Section \ref{Sec creatannih}
we focus on the set of stargenfunctions expressed with generalized Laguerre
functions and investigate how they transform by star-multiplying creation
and annihilation operators to generic complex powers. In Section \ref{Sec
conclusion} we summarize the results and briefly discuss some problems of
using these stargenfunctions to further construct valid solutions to
Vasiliev's equations.

\section{Solve for the stargenfunctions \label{Sec solveeigen}}

In order to solve (\ref{originaleq}), we first use the lemmas 
\begin{subequations}
\begin{eqnarray}
Y_{\underline{\alpha }}\star f\left( Y\right) &=&Y_{\underline{\alpha }%
}f\left( Y\right) +i\frac{\partial }{\partial Y^{\underline{\alpha }}}%
f\left( Y\right) \text{ ,} \\
f\left( Y\right) \star Y_{\underline{\alpha }} &=&Y_{\underline{\alpha }%
}f\left( Y\right) -i\frac{\partial }{\partial Y^{\underline{\alpha }}}%
f\left( Y\right)
\end{eqnarray}%
and (\ref{alinY}) to convert the star-products with number operators into
derivatives w.r.t.\ the creation and annihilation operators, which converts (%
\ref{originaleq}) into 
\end{subequations}
\begin{subequations}
\begin{eqnarray}
\left( a^{+}a^{-}+\frac{1}{2}a^{+}\frac{\partial }{\partial a^{+}}-\frac{1}{2%
}a^{-}\frac{\partial }{\partial a^{-}}-\frac{1}{4}\frac{\partial ^{2}}{%
\partial a^{+}\partial a^{-}}\right) f_{\lambda _{L}|\lambda _{R}}\left(
a^{+},a^{-}\right) &=&\lambda _{L}f_{\lambda _{L}|\lambda _{R}}\left(
a^{+},a^{-}\right) \text{ ,}  \label{eqL} \\
\left( a^{+}a^{-}-\frac{1}{2}a^{+}\frac{\partial }{\partial a^{+}}+\frac{1}{2%
}a^{-}\frac{\partial }{\partial a^{-}}-\frac{1}{4}\frac{\partial ^{2}}{%
\partial a^{+}\partial a^{-}}\right) f_{\lambda _{L}|\lambda _{R}}\left(
a^{+},a^{-}\right) &=&\lambda _{R}f_{\lambda _{L}|\lambda _{R}}\left(
a^{+},a^{-}\right) \text{ ,}  \label{eqR}
\end{eqnarray}%
or equivalently by taking the sum and the difference: 
\end{subequations}
\begin{eqnarray}
\left( a^{+}\frac{\partial }{\partial a^{+}}-a^{-}\frac{\partial }{\partial
a^{-}}\right) f_{\lambda _{L}|\lambda _{R}}\left( a^{+},a^{-}\right)
&=&\left( \lambda _{L}-\lambda _{R}\right) f_{\lambda _{L}|\lambda
_{R}}\left( a^{+},a^{-}\right) \text{ ,}  \label{eqdiff} \\
\left( 2a^{+}a^{-}-\frac{1}{2}\frac{\partial ^{2}}{\partial a^{+}\partial
a^{-}}\right) f_{\lambda _{L}|\lambda _{R}}\left( a^{+},a^{-}\right)
&=&\left( \lambda _{L}+\lambda _{R}\right) f_{\lambda _{L}|\lambda
_{R}}\left( a^{+},a^{-}\right) \text{ .}  \label{eqsum}
\end{eqnarray}%
The solution to (\ref{eqdiff}) can be written as 
\begin{equation}
f_{\lambda _{L}|\lambda _{R}}^{+}\left( a^{+},a^{-}\right) =\left(
a^{+}\right) ^{\lambda _{L}-\lambda _{R}}g_{\lambda _{L}|\lambda
_{R}}^{+}\left( a^{+}a^{-}\right) \text{ \ \ or \ \ }f_{\lambda _{L}|\lambda
_{R}}^{-}\left( a^{+},a^{-}\right) =\left( a^{-}\right) ^{\lambda
_{R}-\lambda _{L}}g_{\lambda _{L}|\lambda _{R}}^{-}\left( a^{+}a^{-}\right) 
\text{ .}  \label{soleqdiff}
\end{equation}%
Note that the two equations stay the same by exchanging $\{a^{+}%
\leftrightarrow a^{-}\ ,\ \ \lambda _{L}\leftrightarrow \lambda _{R}\}$,
corresponding to the exchange of the solutions $f^{+}$ and $f^{-}$ above. By
substituting (\ref{soleqdiff}) into (\ref{eqsum}) we get 
\begin{equation}
2wg_{\lambda _{L}|\lambda _{R}}^{\pm }\left( w\right) -\frac{1}{2}\left( \pm
\lambda _{L}\mp \lambda _{R}+1\right) g_{\lambda _{L}|\lambda _{R}}^{\pm
\prime }\left( w\right) -\frac{1}{2}wg_{\lambda _{L}|\lambda _{R}}^{\pm
\prime \prime }\left( w\right) =\left( \lambda _{L}+\lambda _{R}\right)
g_{\lambda _{L}|\lambda _{R}}^{\pm }\left( w\right) \text{ .}
\label{eqsumtrans}
\end{equation}%
For generic choices of the stargenvalues, solutions to (\ref{eqsumtrans})
can be written as 
\begin{eqnarray}
g_{\lambda _{L}|\lambda _{R}}^{+}\left( w\right) &=&\mathrm{C}%
_{1}e^{-2w}L_{\lambda _{R}-\frac{1}{2}}^{\lambda _{L}-\lambda _{R}}\left(
4w\right) +\mathrm{C}_{2}w^{\lambda _{R}-\lambda _{L}}e^{-2w}L_{\lambda _{L}-%
\frac{1}{2}}^{\lambda _{R}-\lambda _{L}}\left( 4w\right) \text{ ,\ \ } 
\notag \\
\text{ \ \ }g_{\lambda _{L}|\lambda _{R}}^{-}\left( w\right) &=&\mathrm{C}%
_{2}e^{-2w}L_{\lambda _{L}-\frac{1}{2}}^{\lambda _{R}-\lambda _{L}}\left(
4w\right) +\mathrm{C}_{1}w^{\lambda _{L}-\lambda _{R}}e^{-2w}L_{\lambda _{R}-%
\frac{1}{2}}^{\lambda _{L}-\lambda _{R}}\left( 4w\right) \text{ ,}
\label{gSoln}
\end{eqnarray}%
where the $L$ is the generalized Laguerre function, and the $\mathrm{C}$'s
are integration constants. Then we can substitute either one of (\ref{gSoln}%
) into (\ref{soleqdiff}), which gives the solution to the original equations 
\begin{equation}
f_{\lambda _{L}|\lambda _{R}}\left( a^{+},a^{-}\right) =\mathrm{C}_{1}\left(
a^{+}\right) ^{\lambda _{L}-\lambda _{R}}e^{-2w}L_{\lambda _{R}-\frac{1}{2}%
}^{\lambda _{L}-\lambda _{R}}\left( 4w\right) +\mathrm{C}_{2}\left(
a^{-}\right) ^{\lambda _{R}-\lambda _{L}}e^{-2w}L_{\lambda _{L}-\frac{1}{2}%
}^{\lambda _{R}-\lambda _{L}}\left( 4w\right) \text{ .}  \label{fsolnLL}
\end{equation}%
Because the Tricomi confluent hypergeometric function 
\begin{eqnarray}
U\left( \frac{1}{2}-\lambda _{R},1+\lambda _{L}-\lambda _{R},4w\right)
&=&\csc \left[ \pi \left( \lambda _{L}-\lambda _{R}\right) \right] \left[
-\cos \left( \pi \lambda _{L}\right) \Gamma \left( \frac{1}{2}+\lambda
_{R}\right) L_{\lambda _{R}-\frac{1}{2}}^{\lambda _{L}-\lambda _{R}}\left(
4w\right) \right.  \notag \\
&&\ \ \ \ \ \ \left. +\left( 4w\right) ^{\lambda _{R}-\lambda _{L}}\cos
\left( \pi \lambda _{R}\right) \Gamma \left( \frac{1}{2}+\lambda _{L}\right)
L_{\lambda _{L}-\frac{1}{2}}^{\lambda _{R}-\lambda _{L}}\left( 4w\right) %
\right] \text{ ,}  \label{ULLrelation}
\end{eqnarray}%
we can alternatively write (\ref{fsolnLL}) as 
\begin{equation}
f_{\lambda _{L}|\lambda _{R}}\left( a^{+},a^{-}\right) =\left( a^{+}\right)
^{\lambda _{L}-\lambda _{R}}e^{-2w}\left[ \mathrm{C}_{3}L_{\lambda _{R}-%
\frac{1}{2}}^{\lambda _{L}-\lambda _{R}}\left( 4w\right) +\mathrm{C}%
_{4}U\left( \frac{1}{2}-\lambda _{R},1+\lambda _{L}-\lambda _{R},4w\right) %
\right] \text{ .}  \label{fsolnLU}
\end{equation}

There exist subtleties that, for special stargenvalues, the two branches of
solutions presented above may degenerate. For example, the two terms in (\ref%
{fsolnLU}) degenerate when $\lambda _{R}+\frac{1}{2}\in \mathbb{Z}$ because
of the factor $\cos \left( \pi \lambda _{R}\right) $ in (\ref{ULLrelation}).
For another example, the two terms in (\ref{fsolnLL}) degenerate when $%
\lambda _{R}-\lambda _{L}\in \mathbb{Z}$, because 
\begin{eqnarray}
L_{\lambda _{L}-\frac{1}{2}}^{\lambda _{R}-\lambda _{L}}\left( 4w\right) &=&%
\frac{\left( 4w\right) ^{\lambda _{L}-\lambda _{R}}}{\left( \frac{1}{2}%
-\lambda _{L}\right) _{\lambda _{L}-\lambda _{R}}}L_{\lambda _{R}-\frac{1}{2}%
}^{\lambda _{L}-\lambda _{R}}\left( 4w\right) \text{ \ \ for \ }\lambda
_{L}-\lambda _{R}\in \mathbb{Z}^{+}\text{ ,}  \notag \\
L_{\lambda _{R}-\frac{1}{2}}^{\lambda _{L}-\lambda _{R}}\left( 4w\right) &=&%
\frac{\left( 4w\right) ^{\lambda _{R}-\lambda _{L}}}{\left( \frac{1}{2}%
-\lambda _{R}\right) _{\lambda _{R}-\lambda _{L}}}L_{\lambda _{L}-\frac{1}{2}%
}^{\lambda _{R}-\lambda _{L}}\left( 4w\right) \text{ \ \ for \ }\lambda
_{R}-\lambda _{L}\in \mathbb{Z}^{+}\text{ ,}  \label{integerdiff}
\end{eqnarray}%
where $\left( \cdot \right) _{\cdot }$ is the Pochhammer symbol.

Note that when both stargenvalues $\lambda _{L}$ and $\lambda _{R}$ are
half-integers, (\ref{fsolnLL}) and (\ref{fsolnLU}) reduce to the situation
discussed in \cite{Iazeolla:2011cb}, and in \cite{Aros:2019pgj} we only
focused on the situation that either stargenvalue is complex with the other
still being a half-integer. This was because we wanted to limit ourselves to
the small closed contour integral representation of the stargenfunctions.
However, in this paper, we let both stargenvalues be complex numbers in
general, and use different integral representations as shown in the next
section.

\section{Integral representation \label{Sec integral}}

To enable star-product computation of special functions, we very often need
to represent them by integrals with integrands where $Y$-coordinates only
appear in exponents, which we shall do in this section. Note that usually
the integrals cannot cover the whole parameter space, and the prescription
of analytic continuation is needed for the rest of the parameter space after
the integrals are done.

We first rewrite the generalized Laguerre function and the Tricomi confluent
hypergeometric function in terms of integrals on the real axis 
\begin{eqnarray}
e^{-2w}L_{\lambda _{R}-\frac{1}{2}}^{\lambda _{L}-\lambda _{R}}\left(
4w\right) &=&\frac{2^{\lambda _{R}-\lambda _{L}}\cos \left( \pi \lambda
_{R}\right) }{\pi }\int_{-1}^{1}e^{2ws}\frac{\left\vert s-1\right\vert
^{\lambda _{L}-\frac{1}{2}}}{\left\vert s+1\right\vert ^{\lambda _{R}+\frac{1%
}{2}}}ds\text{ ,}  \label{realintL} \\
e^{-2w}U\left( \frac{1}{2}-\lambda _{R},1+\lambda _{L}-\lambda
_{R},4w\right) &=&\frac{2^{\lambda _{R}-\lambda _{L}}}{\Gamma \left( \frac{1%
}{2}-\lambda _{R}\right) }\int_{-\infty }^{-1}e^{2ws}\frac{\left\vert
s-1\right\vert ^{\lambda _{L}-\frac{1}{2}}}{\left\vert s+1\right\vert
^{\lambda _{R}+\frac{1}{2}}}ds\text{ .}  \label{realintU}
\end{eqnarray}%
To be precise, (\ref{realintL}) holds when $\mathrm{Re}\left( \lambda
_{R}\right) <\frac{1}{2}$ and $\mathrm{Re}\left( \lambda _{L}\right) >-\frac{%
1}{2}$, and (\ref{realintU}) holds when $\mathrm{Re}\left( \lambda
_{L}\right) >-\frac{1}{2}$ and $\mathrm{Re}\left( w\right) >0$.

We can further convert them into contour integrals on the complex $s$-plane.

\begin{figure}[h]
\caption{}
\label{Cminfm1}\centering
\includegraphics[scale=0.5]{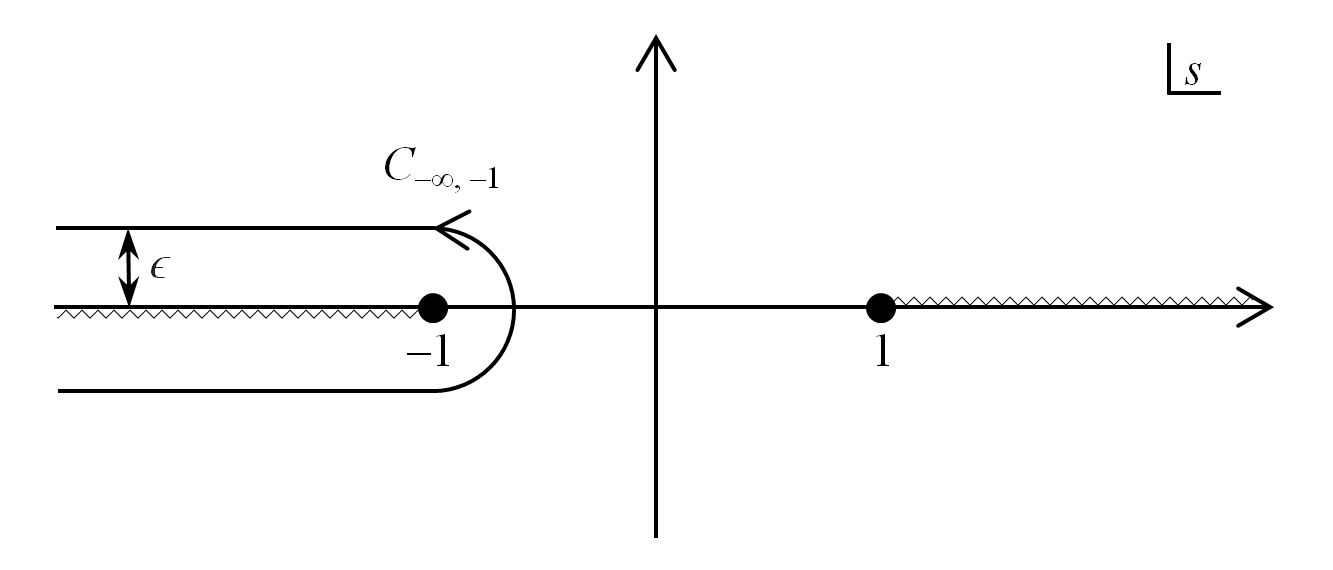}
\end{figure}

Let us define the contour integral (Figure \ref{Cminfm1}) 
\begin{equation}
\oint_{C_{-\infty ,-1}}=\int_{-\infty -i\epsilon }^{-1-i\epsilon
}+\int_{-1-i\epsilon }^{-1+i\epsilon }+\int_{-1+i\epsilon }^{-\infty
+i\epsilon }\text{ ,}
\end{equation}%
where $\epsilon $ is an infinitesimally small positive number, and $%
\int_{-1-i\epsilon }^{-1+i\epsilon }$ is a half-circle surrounding the right
side of $-1$ on the complex plane. We adopt the branch cuts that are
consistent with 
\begin{equation}
\mathrm{Ln}\left( 1-s\right) \text{\ \ \ and\ \ \ }\mathrm{Ln}\left(
1+s\right) \text{ ,}
\end{equation}%
where $\mathrm{Ln}$ uses the principal branch of the phase angle $(-\pi ,\pi
]$ of its parameter. With this convention we can prove that 
\begin{eqnarray}
&&\int_{-\infty -i\epsilon }^{-1-i\epsilon }e^{2ws}\frac{\left( 1-s\right)
^{\lambda _{L}-\frac{1}{2}}}{\left( 1+s\right) ^{\lambda _{R}+\frac{1}{2}}}ds%
\overset{\epsilon \rightarrow 0}{=}\int_{-\infty }^{-1}e^{2ws}\frac{%
\left\vert s-1\right\vert ^{\lambda _{L}-\frac{1}{2}}}{\left\vert
s+1\right\vert ^{\lambda _{R}+\frac{1}{2}}e^{-i\pi \left( \lambda _{R}+\frac{%
1}{2}\right) }}ds\text{ ,}  \notag \\
&&\int_{-1+i\epsilon }^{-\infty +i\epsilon }e^{2ws}\frac{\left( 1-s\right)
^{\lambda _{L}-\frac{1}{2}}}{\left( 1+s\right) ^{\lambda _{R}+\frac{1}{2}}}ds%
\overset{\epsilon \rightarrow 0}{=}\int_{-1}^{-\infty }e^{2ws}\frac{%
\left\vert s-1\right\vert ^{\lambda _{L}-\frac{1}{2}}}{\left\vert
s+1\right\vert ^{\lambda _{R}+\frac{1}{2}}e^{i\pi \left( \lambda _{R}+\frac{1%
}{2}\right) }}ds\text{ .}  \label{minfm1c}
\end{eqnarray}%
Furthermore, for the half-circle integral we can do a change of the variable 
$s=\epsilon e^{i\theta }-1$, then%
\begin{equation}
\int_{-1-i\epsilon }^{-1+i\epsilon }e^{2ws}\frac{\left( 1-s\right) ^{\lambda
_{L}-\frac{1}{2}}}{\left( 1+s\right) ^{\lambda _{R}+\frac{1}{2}}}ds=\int_{-%
\frac{\pi }{2}}^{\frac{\pi }{2}}e^{2w\left( \epsilon e^{i\theta }-1\right) }%
\frac{\left( 2-\epsilon e^{i\theta }\right) ^{\lambda _{L}-\frac{1}{2}}}{%
\left( \epsilon e^{i\theta }\right) ^{\lambda _{R}+\frac{1}{2}}}i\epsilon
e^{i\theta }d\theta \text{ ,}
\end{equation}%
which, by counting the power of $\epsilon $, obviously vanishes in the limit
\ $\epsilon \rightarrow 0$ when $\mathrm{Re}\left( \lambda _{R}\right) <%
\frac{1}{2}$. Therefore, using (\ref{minfm1c}) and (\ref{realintU}) we can
derive 
\begin{eqnarray}
&&\oint_{C_{-\infty ,-1}}e^{2ws}\frac{\left( 1-s\right) ^{\lambda _{L}-\frac{%
1}{2}}}{\left( 1+s\right) ^{\lambda _{R}+\frac{1}{2}}}ds  \notag \\
&=&2^{1+\lambda _{L}-\lambda _{R}}i\cos \left( \pi \lambda _{R}\right)
\Gamma \left( \frac{1}{2}-\lambda _{R}\right) e^{-2w}U\left( \frac{1}{2}%
-\lambda _{R},1+\lambda _{L}-\lambda _{R},4w\right) \text{ .}
\end{eqnarray}

\begin{figure}[h]
\caption{}
\label{Cminfp1}\centering
\includegraphics[scale=0.5]{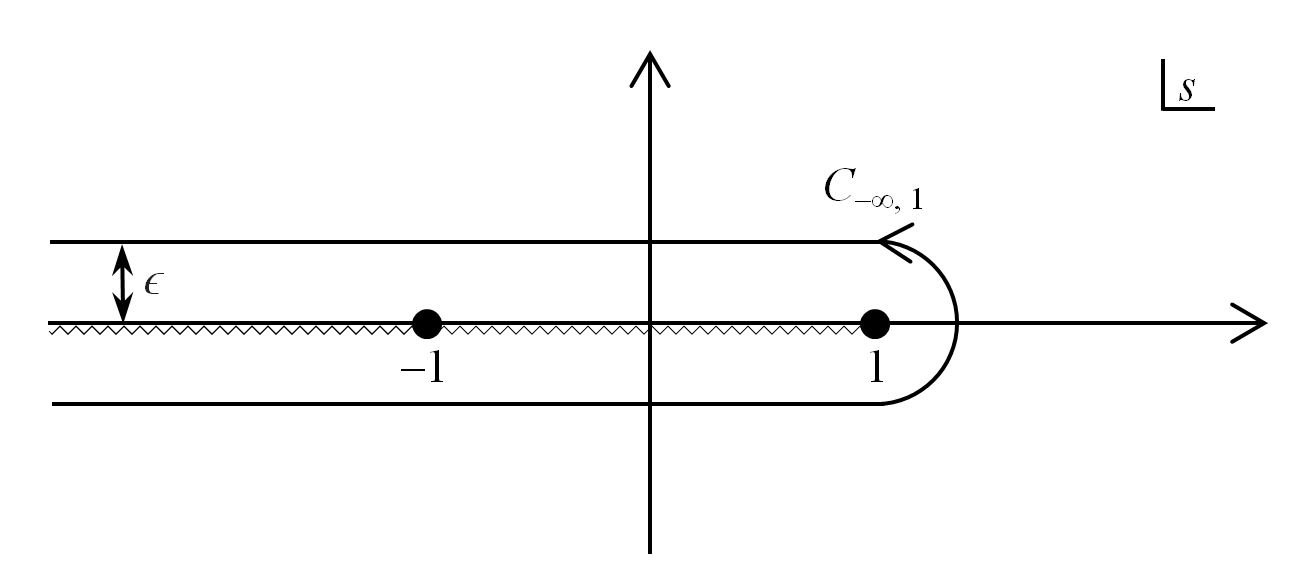}
\end{figure}

Let us also define the contour integral (Figure \ref{Cminfp1}) 
\begin{equation}
\oint_{C_{-\infty ,1}}=\int_{-\infty -i\epsilon }^{1-i\epsilon
}+\int_{1-i\epsilon }^{1+i\epsilon }+\int_{1+i\epsilon }^{-\infty +i\epsilon
}\text{ ,}
\end{equation}%
where $\int_{1-i\epsilon }^{1+i\epsilon }$ is a half-circle surrounding the
right side of $1$ on the complex plane. Here we adopt the branch cuts that
are consistent with 
\begin{equation}
\mathrm{Ln}\left( s-1\right) \text{ \ \ and \ \ }\mathrm{Ln}\left(
s+1\right) \text{ .}
\end{equation}%
Then we can derive 
\begin{eqnarray}
&&\int_{-\infty -i\epsilon }^{-1-i\epsilon }e^{2ws}\frac{\left( s-1\right)
^{\lambda _{L}-\frac{1}{2}}}{\left( s+1\right) ^{\lambda _{R}+\frac{1}{2}}}ds%
\overset{\epsilon \rightarrow 0}{=}\int_{-\infty }^{-1}e^{2ws}\frac{%
\left\vert s-1\right\vert ^{\lambda _{L}-\frac{1}{2}}e^{-i\pi \left( \lambda
_{L}-\frac{1}{2}\right) }}{\left\vert s+1\right\vert ^{\lambda _{R}+\frac{1}{%
2}}e^{-i\pi \left( \lambda _{R}+\frac{1}{2}\right) }}ds\text{ ,}  \notag \\
&&\int_{-1-i\epsilon }^{1-i\epsilon }e^{2ws}\frac{\left( s-1\right)
^{\lambda _{L}-\frac{1}{2}}}{\left( s+1\right) ^{\lambda _{R}+\frac{1}{2}}}ds%
\overset{\epsilon \rightarrow 0}{=}\int_{-1}^{1}e^{2ws}\frac{\left\vert
s-1\right\vert ^{\lambda _{L}-\frac{1}{2}}e^{-i\pi \left( \lambda _{L}-\frac{%
1}{2}\right) }}{\left\vert s+1\right\vert ^{\lambda _{R}+\frac{1}{2}}}ds%
\text{ ,}  \notag \\
&&\int_{-1-i\epsilon }^{1-i\epsilon }e^{2ws}\frac{\left( s-1\right)
^{\lambda _{L}-\frac{1}{2}}}{\left( s+1\right) ^{\lambda _{R}+\frac{1}{2}}}ds%
\overset{\epsilon \rightarrow 0}{=}\int_{-1}^{1}e^{2ws}\frac{\left\vert
s-1\right\vert ^{\lambda _{L}-\frac{1}{2}}e^{i\pi \left( \lambda _{L}-\frac{1%
}{2}\right) }}{\left\vert s+1\right\vert ^{\lambda _{R}+\frac{1}{2}}}ds\text{
,}  \notag \\
&&\int_{-1+i\epsilon }^{-\infty +i\epsilon }e^{2ws}\frac{\left( s-1\right)
^{\lambda _{L}-\frac{1}{2}}}{\left( s+1\right) ^{\lambda _{R}+\frac{1}{2}}}ds%
\overset{\epsilon \rightarrow 0}{=}\int_{-1}^{-\infty }e^{2ws}\frac{%
\left\vert s-1\right\vert ^{\lambda _{L}-\frac{1}{2}}e^{i\pi \left( \lambda
_{L}-\frac{1}{2}\right) }}{\left\vert s+1\right\vert ^{\lambda _{R}+\frac{1}{%
2}}e^{i\pi \left( \lambda _{R}+\frac{1}{2}\right) }}ds\text{ .}
\label{minf1c}
\end{eqnarray}%
Again, we can see with a change of the variable $s=\epsilon e^{i\theta }+1$
that the half-circle integral 
\begin{equation}
\int_{1-i\epsilon }^{1+i\epsilon }e^{2ws}\frac{\left( s-1\right) ^{\lambda
_{L}-\frac{1}{2}}}{\left( s+1\right) ^{\lambda _{R}+\frac{1}{2}}}ds=\int_{-%
\frac{\pi }{2}}^{\frac{\pi }{2}}e^{2w\left( \epsilon e^{i\theta }+1\right) }%
\frac{\left( \epsilon e^{i\theta }\right) ^{\lambda _{L}-\frac{1}{2}}}{%
\left( \epsilon e^{i\theta }+2\right) ^{\lambda _{R}+\frac{1}{2}}}i\epsilon
e^{i\theta }d\theta
\end{equation}%
vanishes in the limit $\epsilon \rightarrow 0$ when $\mathrm{Re}\left(
\lambda _{L}\right) >-\frac{1}{2}$. Therefore, combining (\ref{minf1c}), (%
\ref{realintL}) and (\ref{realintU}) we can derive 
\begin{eqnarray}
\oint_{C_{-\infty ,1}}e^{2ws}\frac{\left( s-1\right) ^{\lambda _{L}-\frac{1}{%
2}}}{\left( s+1\right) ^{\lambda _{R}+\frac{1}{2}}}ds &=&2^{1+\lambda
_{L}-\lambda _{R}}\pi i\cos \left( \pi \lambda _{L}\right) \sec \left( \pi
\lambda _{R}\right) e^{-2w}L_{\lambda _{R}-\frac{1}{2}}^{\lambda
_{L}-\lambda _{R}}\left( 4w\right)  \notag \\
&&+2^{1+\lambda _{L}-\lambda _{R}}i\sin \left[ \pi \left( \lambda
_{L}-\lambda _{R}\right) \right] \Gamma \left( \frac{1}{2}-\lambda
_{R}\right) e^{-2w}U\left( \frac{1}{2}-\lambda _{R},1+\lambda _{L}-\lambda
_{R},4w\right) \text{ .}  \notag \\
&&
\end{eqnarray}%
The second term on the r.h.s.\ is the result of $\int_{-\infty -i\epsilon
}^{-1-i\epsilon }+\int_{-1+i\epsilon }^{-\infty +i\epsilon }$, and it
vanishes when $\lambda _{L}-\lambda _{R}\in \mathbb{Z}$, which can be
explained by the cancellation of the branch cuts where $s<-1$ on the
real-axis.

\begin{figure}[h]
\caption{}
\label{Cm1p1}\centering
\includegraphics[scale=0.5]{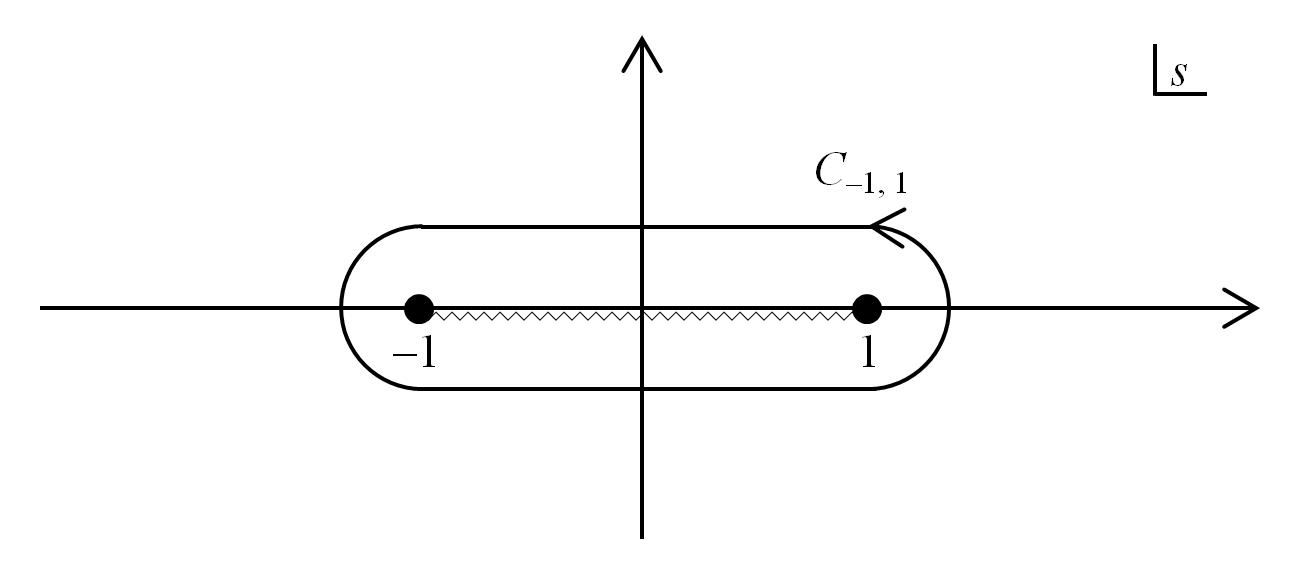}
\end{figure}

To summarize, we can rewrite the solution (\ref{fsolnLU}) as 
\begin{equation}
f_{\lambda _{L}|\lambda _{R}}\left( a^{+},a^{-}\right) =\left( a^{+}\right)
^{\lambda _{L}-\lambda _{R}}\left[ \mathrm{C}_{5}\oint_{C_{-\infty
,1}}e^{2ws}\frac{\left( s-1\right) ^{\lambda _{L}-\frac{1}{2}}}{\left(
s+1\right) ^{\lambda _{R}+\frac{1}{2}}}ds+\mathrm{C}_{6}\oint_{C_{-\infty
,-1}}e^{2ws}\frac{\left( 1-s\right) ^{\lambda _{L}-\frac{1}{2}}}{\left(
1+s\right) ^{\lambda _{R}+\frac{1}{2}}}\right] \text{ ,}
\label{sum two contours}
\end{equation}%
and it is important to notice that the convergence of the above contour
integrals requires only Re$\left( w\right) >0$ and has no restrictions on $%
\lambda _{L,R}$. In the diagonal case $\lambda _{L}=\lambda _{R}$, due to
the cancellation between two branch cuts, $\oint_{C_{-\infty ,1}}$ can be
replaced by $\oint_{C_{-1,1}}$, which is a contour integral surrounding the
section of the real axis between $-1$ and $1$ (Figure \ref{Cm1p1}).

At the end of this section, let us briefly compare the contours in this
paper and the previous ones in \cite{Iazeolla:2011cb,Aros:2019pgj}. The
stargenfunctions in \cite{Aros:2019pgj} (or \cite{Iazeolla:2011cb}) are just
special cases in Section \ref{Sec solveeigen} with one (or two) of the left
and right stargenvalues set to half-integers. Therefore, we should be able
to reproduce these special cases also from the contour integrals.\footnote{%
See \cite{Iazeolla:2008ix} for relevant discussions in details.} Take the
first term on the r.h.s.\ of (\ref{sum two contours}) for example, if we set 
$\lambda _{L}=\lambda _{R}\in \mathbb{Z+}\frac{1}{2}$, all brunch cuts can
be dropped, then the contour reduces to two circles around the points $s=\pm
1$, one of which has zero residue, and thus the integral reproduces
stargenfunctions with Laguerre polynomials as in \cite{Iazeolla:2011cb}.
Take the second term for another example, if we set $\lambda _{R}\mathbb{+}%
\frac{1}{2}\in \mathbb{Z}^{+}$, the left-side brunch cut in Figure \ref%
{Cminfm1} goes away, then the contour reduces to a circle around $s=-1$, and
thus by keeping $\lambda _{L}\in $ $\mathbb{C}$ with the other brunch cut,
the integral reproduces stargenfunctions with generalized Laguerre functions
as in \cite{Aros:2019pgj}.

\section{Creation and annihilation operators \label{Sec creatannih}}

The set of stargenfunctions (\ref{fsolnLL}) is the most direct
generalization from \cite{Iazeolla:2011cb,Aros:2019pgj}. In this section, we
investigate how they change by star-multiplying creation and/or annihilation
operators to complex powers. Let us call the stargenfunctions represented by
the two terms of (\ref{fsolnLL}), respectively, the \textquotedblleft
plus\textquotedblright\ and the \textquotedblleft minus\textquotedblright\
branches. For convenience of star-product computation, we write the two
branches by the integral representation (\ref{realintL}), i.e. \footnote{%
In this section, we adopt real-axis integrals instead of contour integrals.
Although the latter has fewer restrictions on its parameters, the former has
fewer subtleties related to branch cuts and hence is more convenient for
computation involving complex powers. For example, the step from (\ref%
{intbeforeresc}) to (\ref{intafterresc}) is easy and clear, thanks to the
fact that every exponent in the integrands has a positive real base and thus
formulas like $\left( a/b\right) ^{z}=a^{z}b^{-z}$ with $z\in \mathbb{C}$
can be safely used.} \ 
\begin{subequations}
\label{fplusminus}
\begin{eqnarray}
f_{\lambda _{L}|\lambda _{R}}^{+}\left( a^{+},a^{-}\right) &=&\left(
a^{+}\right) ^{\lambda _{L}-\lambda _{R}}\int_{-1}^{1}e^{2ws}\frac{\left(
1-s\right) ^{\lambda _{L}-\frac{1}{2}}}{\left( 1+s\right) ^{\lambda _{R}+%
\frac{1}{2}}}ds\text{ ,} \\
f_{\lambda _{L}|\lambda _{R}}^{-}\left( a^{+},a^{-}\right) &=&\left(
a^{-}\right) ^{\lambda _{R}-\lambda _{L}}\int_{-1}^{1}e^{2ws}\frac{\left(
1-s\right) ^{\lambda _{R}-\frac{1}{2}}}{\left( 1+s\right) ^{\lambda _{L}+%
\frac{1}{2}}}ds\text{ ,}
\end{eqnarray}%
where the constant overall factors have been dropped hereafter for
simplicity. Note that the two branches are in general different, although
they have the same stargenvalues, and that the two degenerate into one when $%
\lambda _{R}-\lambda _{L}\in \mathbb{Z}$ as already discussed around (\ref%
{integerdiff}). In this section, we will investigate how these functions (as
if they were eigenstates of the number operator in quantum mechanics) can be
related by the creation and annihilation operators.

\subsection{The plus and minus branches}

We start from the diagonal case $\lambda _{L}=\lambda _{R}=\lambda $ 
\end{subequations}
\begin{equation}
f_{\lambda |\lambda }\left( a^{+},a^{-}\right) =\int_{-1}^{1}e^{2ws}\frac{%
\left( 1-s\right) ^{\lambda -\frac{1}{2}}}{\left( 1+s\right) ^{\lambda +%
\frac{1}{2}}}ds\text{ .}  \label{fllint}
\end{equation}%
We would like to show that by doing a left or a right star-product with $%
a^{\pm }$ to a complex power can shift the stargenvalue accordingly, and to
this end we first express $a^{\pm }$ with the power $p$ using the Mellin
transform (as done in \cite{Aros:2019pgj}) 
\begin{equation}
\left( a^{\pm }\right) ^{p}=\int_{0}^{+\infty }\frac{\tau ^{-p-1}}{\Gamma
\left( -p\right) }e^{-\tau a^{\pm }}d\tau \text{ ,}  \label{Mellintransf}
\end{equation}%
so that we can pick out the factor $e^{-\tau a^{\pm }}$ for the convenience
of doing star-products. \footnote{%
We impose the restrictions $\left\vert \mathrm{Re}\left( \lambda
_{(L,R)}\right) \right\vert <\frac{1}{2}$, $\mathrm{Re}\left( a^{\pm
}\right) >0$ and $\mathrm{Re}\left( p\right) <0$ for convergence of the
above integrals, but after we have finished the star-product computation and
completed the integrals, we obtain back formulas like $\left( a^{\pm
}\right) ^{p}$ and the generalized Laguerre functions, then by analytic
continuation we can lift these restrictions.} Then by using (\ref{fplusminus}%
), (\ref{Mellintransf}) and the following set of star-product results 
\begin{subequations}
\label{tauasw}
\begin{eqnarray}
e^{-\tau a^{+}}\star e^{2sw} &=&e^{2sw-\tau \left( 1-s\right) a^{+}}\text{ ,}
\\
e^{-\tau a^{-}}\star e^{2sw} &=&e^{2sw-\tau \left( 1+s\right) a^{-}}\text{ ,}
\\
e^{2sw}\star e^{-\tau a^{+}} &=&e^{2sw-\tau \left( 1+s\right) a^{+}}\text{ ,}
\\
e^{2sw}\star e^{-\tau a^{-}} &=&e^{2sw-\tau \left( 1-s\right) a^{-}}\text{ ,}
\end{eqnarray}%
we can easily prove that 
\end{subequations}
\begin{subequations}
\label{eigenshift}
\begin{eqnarray}
f_{\lambda _{L}|\lambda _{R}}^{+} &=&\left( a^{+}\right) ^{\lambda
_{L}-\lambda _{R}}\star f_{\lambda _{R}|\lambda _{R}}=f_{\lambda
_{L}|\lambda _{L}}\star \left( a^{+}\right) ^{\lambda _{L}-\lambda _{R}}%
\text{ ,}  \label{eigenshiftcr} \\
f_{\lambda _{L}|\lambda _{R}}^{-} &=&\left( a^{-}\right) ^{\lambda
_{R}-\lambda _{L}}\star f_{\lambda _{R}|\lambda _{R}}=f_{\lambda
_{L}|\lambda _{L}}\star \left( a^{-}\right) ^{\lambda _{R}-\lambda _{L}}%
\text{ .}
\end{eqnarray}%
Thus the shift of the left (right) stargenvalue is indeed equal to the power
of the creation (annihilation) operator acting from the left (right); or
equal to the opposite power if it is the annihilation (creation) operator.
Furthermore (\ref{eigenshift}) shows that, starting from a diagonal
stargenfunction, the creation / annihilation operator always brings it to
the plus / minus branch, no matter whether the star-product is done from the
left or from the right, and any non-diagonal stargenfunction of the two
branches can be reached in this way, up to an overall constant, from a
diagonal stargenfunction.

Naively we can further use $\left( a^{\pm }\right) ^{p_{1}}\star \left(
a^{\pm }\right) ^{p_{2}}=\left( a^{\pm }\right) ^{p_{1}+p_{2}}$ to conclude
that the plus / minus branch is always closed under the star-multiplications
with creation / annihilation operators to generic complex powers, but here
we encounter some subtlety about associativity.

Star-products are not well-defined unless associativity holds, and
associativity is bound to work if all the formulas to be star-multiplied are
Taylor-expandable in the $Y$-coordinates. However, $a^{\pm }$ to complex
powers cannot be Taylor-expanded. To circumvent the problem, we need to use
integral representations, so that all the $a^{\pm }$'s stay in exponents,
which makes the formulas Taylor-expandable. In other words, the integrals
that we have used in this paper are not only technical tools to simplify the
computation, but also very often the prescription to guarantee
associativity. To ensure associativity during the computation, we need to
keep in mind that all star-products should be computed before doing the
integrals.\footnote{%
See also \cite{Aros:2019pgj,DeFilippi:2019jqq} for relevant discussions.}

For example, in order to show

\end{subequations}
\begin{equation}
\left( a^{+}\right) ^{p_{1}}\star \left( a^{+}\right) ^{p_{2}}\star
f_{\lambda |\lambda }=f_{\lambda +p_{1}+p_{2}|\lambda }^{+}\text{ ,}
\end{equation}%
we need to express all the factors by the integrals:%
\begin{eqnarray}
&&\left( a^{+}\right) ^{p_{1}}\star \left( a^{+}\right) ^{p_{2}}\star
f_{\lambda |\lambda }  \notag \\
&=&\int_{0}^{+\infty }d\tau _{1}\int_{0}^{+\infty }d\tau _{2}\int_{-1}^{1}ds%
\frac{\tau _{1}^{-p_{1}-1}}{\Gamma \left( -p_{1}\right) }\frac{\tau
_{2}^{-p_{2}-1}}{\Gamma \left( -p_{2}\right) }\frac{\left( 1-s\right)
^{\lambda -\frac{1}{2}}}{\left( 1+s\right) ^{\lambda +\frac{1}{2}}}e^{-\tau
_{1}a^{+}}\star e^{-\tau _{2}a^{+}}\star e^{2ws}\text{ ,}
\end{eqnarray}%
then we should do the integrals only after finishing the star-products
within the integrand. By doing the star-products, the above formula equals%
\begin{equation}
\int_{0}^{+\infty }d\tau _{1}\int_{0}^{+\infty }d\tau _{2}\int_{-1}^{1}ds%
\frac{\tau _{1}^{-p_{1}-1}}{\Gamma \left( -p_{1}\right) }\frac{\tau
_{2}^{-p_{2}-1}}{\Gamma \left( -p_{2}\right) }\frac{\left( 1-s\right)
^{\lambda -\frac{1}{2}}}{\left( 1+s\right) ^{\lambda +\frac{1}{2}}}%
e^{2sw-\left( \tau _{1}+\tau _{2}\right) \left( 1-s\right) a^{+}}\text{ ,}
\label{intbeforeresc}
\end{equation}%
and by rescaling $\tau _{1,2}$ it further equals%
\begin{equation}
\int_{0}^{+\infty }d\tau _{1}\int_{0}^{+\infty }d\tau _{2}\int_{-1}^{1}ds%
\frac{\tau _{1}^{-p_{1}-1}}{\Gamma \left( -p_{1}\right) }\frac{\tau
_{2}^{-p_{2}-1}}{\Gamma \left( -p_{2}\right) }\frac{\left( 1-s\right)
^{\lambda +p_{1}+p_{2}-\frac{1}{2}}}{\left( 1+s\right) ^{\lambda +\frac{1}{2}%
}}e^{2sw-\left( \tau _{1}+\tau _{2}\right) a^{+}}\text{ .}
\label{intafterresc}
\end{equation}%
In this way all star-products have been done between exponential functions
of $Y$, thus the problem of associativity has been circumvented, and we can
then do the integrals in (\ref{intafterresc}), which gives $f_{\lambda
+p_{1}+p_{2}|\lambda }^{+}$ as expected. Using this method we can prove that
indeed the plus (minus) branch is closed by star-multiplications with
arbitrarily many factors of creation (annihilation) operators.

\subsection{Mixing the branches}

Now naturally we want to ask the question: if both the creation and the
annihilation operators together act on a stargenfunction, then what happens?

The first thing to check is whether the result is still a stargenfunction.
As a test, we now examine whether 
\begin{equation}
\left( a^{\mp }\right) ^{\lambda ^{\mp }}\star \left( a^{\pm }\right)
^{\lambda ^{\pm }}\star f_{\lambda |\lambda }  \label{aaftest}
\end{equation}%
is a stargenfunction of $w$.

Using the lemma%
\begin{equation}
e^{-\tau ^{\mp }a^{\mp }}\star e^{-\tau ^{\pm }a^{\pm }}\star
e^{2sw}=e^{-\tau ^{-}a^{-}\left( 1+s\right) -\tau ^{+}a^{+}\left( 1-s\right)
+\frac{1}{2}\left( \pm 1-s\right) \tau ^{-}\tau ^{+}+2sw}\text{ ,}
\end{equation}%
we obtain%
\begin{eqnarray}
&&\left( a^{\mp }\right) ^{\lambda ^{\mp }}\star \left( a^{\pm }\right)
^{\lambda ^{\pm }}\star f_{\lambda |\lambda }  \notag \\
&=&\int_{-1}^{1}ds\int_{0}^{+\infty }d\tau ^{+}\int_{0}^{+\infty }d\tau ^{-}%
\frac{\left( 1-s\right) ^{\lambda -\frac{1}{2}}}{\left( 1+s\right) ^{\lambda
+\frac{1}{2}}}\frac{\left( \tau ^{+}\right) ^{-\lambda ^{+}-1}}{\Gamma
\left( -\lambda ^{+}\right) }\frac{\left( \tau ^{-}\right) ^{-\lambda ^{-}-1}%
}{\Gamma \left( -\lambda ^{-}\right) }e^{2sw-\tau ^{-}a^{-}\left( 1+s\right)
-\tau ^{+}a^{+}\left( 1-s\right) +\frac{1}{2}\left( \pm 1-s\right) \tau
^{-}\tau ^{+}}  \notag \\
&=&\int_{-1}^{1}ds\int_{0}^{+\infty }d\tau ^{+}\int_{0}^{+\infty }d\tau ^{-}%
\frac{\left( 1-s\right) ^{\lambda +\lambda ^{+}-\frac{1}{2}}}{\left(
1+s\right) ^{\lambda -\lambda ^{-}+\frac{1}{2}}}\frac{\left( \tau
^{+}\right) ^{-\lambda ^{+}-1}}{\Gamma \left( -\lambda ^{+}\right) }\frac{%
\left( \tau ^{-}\right) ^{-\lambda ^{-}-1}}{\Gamma \left( -\lambda
^{-}\right) }e^{2sw-\tau ^{-}a^{-}-\tau ^{+}a^{+}+\frac{\pm 1-s}{2\left(
1-s^{2}\right) }\tau ^{-}\tau ^{+}}\text{ ,}
\end{eqnarray}%
where in the last step we have simplified the formula by rescaling the
integration variables $\tau ^{\pm }$. We denote%
\begin{eqnarray}
&&g\left( \lambda ,\lambda ^{+},\lambda ^{-},c\right)  \notag \\
&=&\int_{-1}^{1}ds\int_{0}^{+\infty }d\tau ^{+}\int_{0}^{+\infty }d\tau ^{-}%
\frac{\left( 1-s\right) ^{\lambda +\lambda ^{+}-\frac{1}{2}}}{\left(
1+s\right) ^{\lambda -\lambda ^{-}+\frac{1}{2}}}\frac{\left( \tau
^{+}\right) ^{-\lambda ^{+}-1}}{\Gamma \left( -\lambda ^{+}\right) }\frac{%
\left( \tau ^{-}\right) ^{-\lambda ^{-}-1}}{\Gamma \left( -\lambda
^{-}\right) }e^{2sw-\tau ^{-}a^{-}-\tau ^{+}a^{+}-\frac{c+s}{2\left(
1-s^{2}\right) }\tau ^{-}\tau ^{+}}\text{ .}  \label{gllc}
\end{eqnarray}%
For convergence, we restrict Re$\left( \lambda \pm \lambda ^{\pm }\right) <%
\frac{1}{2}$, Re$\left( \lambda ^{\pm }\right) <0$ and Re$\left( c\right) >1$%
, and (\ref{aaftest}) is restored by analytically continuing $g\left(
\lambda ,\lambda ^{+},\lambda ^{-},c\right) $ and setting $c=\mp 1$. We can
prove that (\ref{aaftest}) is still a stargenfunction of $w$, if we can
prove 
\begin{eqnarray}
w\star g\left( \lambda ,\lambda ^{+},\lambda ^{-},c\right) &=&\left( \lambda
+\lambda ^{+}-\lambda ^{-}\right) g\left( \lambda ,\lambda ^{+},\lambda
^{-},c\right) \text{ ,}  \label{stargentoprove} \\
g\left( \lambda ,\lambda ^{+},\lambda ^{-},c\right) \star w &=&\lambda
g\left( \lambda ,\lambda ^{+},\lambda ^{-},c\right) \text{ .}
\label{stargentoproveright}
\end{eqnarray}%
To this end, there is no need to perform the integrals of (\ref{gllc}), we
only need to use its integrand to read out the stargenvalues.

In the same way as deriving (\ref{eqL}), we convert the $w\star $ on the
l.h.s.\ of (\ref{stargentoprove}) into derivatives w.r.t.\ $a^{\pm }$ and
hence obtain

\begin{eqnarray}
&&w\star e^{2sw-\tau ^{-}a^{-}-\tau ^{+}a^{+}}  \notag \\
&=&\left[ -\frac{s}{2}+w\left( 1-s^{2}\right) -\frac{1}{2}\tau
^{+}a^{+}\left( 1-s\right) +\frac{1}{2}\tau ^{-}a^{-}\left( 1+s\right) -%
\frac{1}{4}\tau ^{+}\tau ^{-}\right] e^{2sw-\tau ^{-}a^{-}-\tau ^{+}a^{+}}%
\text{ ,}
\end{eqnarray}%
which leads to 
\begin{eqnarray}
&&w\star g\left( \lambda ,\lambda ^{+},\lambda ^{-},c\right)  \notag \\
&=&\int_{-1}^{1}ds\int_{0}^{+\infty }d\tau ^{+}\int_{0}^{+\infty }d\tau ^{-}%
\frac{\left( 1-s\right) ^{\lambda +\lambda ^{+}-\frac{1}{2}}}{\left(
1+s\right) ^{\lambda -\lambda ^{-}+\frac{1}{2}}}\frac{\left( \tau
^{+}\right) ^{-\lambda ^{+}-1}}{\Gamma \left( -\lambda ^{+}\right) }\frac{%
\left( \tau ^{-}\right) ^{-\lambda ^{-}-1}}{\Gamma \left( -\lambda
^{-}\right) }e^{2sw-\tau ^{-}a^{-}-\tau ^{+}a^{+}-\frac{c+s}{2\left(
1-s^{2}\right) }\tau ^{-}\tau ^{+}}  \notag \\
&&\ \ \ \ \left[ -\frac{s}{2}+w\left( 1-s^{2}\right) -\frac{1}{2}\tau
^{+}a^{+}\left( 1-s\right) +\frac{1}{2}\tau ^{-}a^{-}\left( 1+s\right) -%
\frac{1}{4}\tau ^{+}\tau ^{-}\right]  \notag
\end{eqnarray}%
\begin{subequations}
\label{fourterms}
\begin{eqnarray}
&=&\int_{-1}^{1}ds\int_{0}^{+\infty }d\tau ^{+}\int_{0}^{+\infty }d\tau ^{-}%
\frac{\left( 1-s\right) ^{\lambda +\lambda ^{+}-\frac{1}{2}}}{\left(
1+s\right) ^{\lambda -\lambda ^{-}+\frac{1}{2}}}\frac{\left( \tau
^{+}\right) ^{-\lambda ^{+}-1}}{\Gamma \left( -\lambda ^{+}\right) }\frac{%
\left( \tau ^{-}\right) ^{-\lambda ^{-}-1}}{\Gamma \left( -\lambda
^{-}\right) }e^{2sw-\tau ^{-}a^{-}-\tau ^{+}a^{+}-\frac{c+s}{2\left(
1-s^{2}\right) }\tau ^{-}\tau ^{+}}  \notag \\
&&\ \ \ \ \left\{ e^{-2sw}\left[ -\frac{s}{2}+\frac{1}{2}\left(
1-s^{2}\right) \frac{\partial }{\partial s}\right] e^{2sw}\right.
\label{terma} \\
&&\ \ \ \ \left. +e^{\tau ^{+}a^{+}}\left[ \frac{1}{2}\tau ^{+}\left(
1-s\right) \frac{\partial }{\partial \tau ^{+}}\right] e^{-\tau
^{+}a^{+}}\right.  \label{termb} \\
&&\ \ \ \ \left. +e^{\tau ^{-}a^{-}}\left[ -\frac{1}{2}\tau ^{-}\left(
1+s\right) \frac{\partial }{\partial \tau ^{-}}\right] e^{-\tau
^{-}a^{-}}\right.  \label{termc} \\
&&\ \ \ \ \left. -\frac{1}{4}\tau ^{+}\tau ^{-}\right\} \text{ .}
\label{termd}
\end{eqnarray}%
We can convert the terms (\ref{terma}) (\ref{termb}) and (\ref{termc}) by
integrations by parts, so that the curly bracket above can be replaced with 
\end{subequations}
\begin{eqnarray*}
&&\ \ \ \left\{ \left[ \lambda +\frac{1}{2}\left( 1+s\right) \lambda ^{+}-%
\frac{1}{2}\left( 1-s\right) \lambda ^{-}+\frac{1+2cs+s^{2}}{4\left(
1-s^{2}\right) }\tau ^{+}\tau ^{-}\right] \right. \  \\
&&\ \ \ \ \left. +\left[ \frac{1}{2}\left( 1-s\right) \lambda ^{+}+\frac{c+s%
}{4\left( 1+s\right) }\tau ^{+}\tau ^{-}\right] \right. \\
&&\ \ \ \ \left. +\left[ -\frac{1}{2}\left( 1+s\right) \lambda ^{-}-\frac{c+s%
}{4\left( 1-s\right) }\tau ^{+}\tau ^{-}\right] \right. \\
&&\ \ \ \ \left. -\frac{1}{4}\tau ^{+}\tau ^{-}\right\} \text{ ,}
\end{eqnarray*}%
which is exactly equal to the factor $\left( \lambda +\lambda ^{+}-\lambda
^{-}\right) $, and thus (\ref{stargentoprove}) is proven. Moreover (\ref%
{stargentoproveright}) can also be proven in the same manner, whose details
we will skip.

In (\ref{aaftest}), we can use (\ref{eigenshift}) to move one or both of the
creation and annihilation factors to the right side, and thus it is clear
that a diagonal stargenfunction of the type (\ref{fsolnLL}), star-multiplied
with two factors respectively of creation and annihilation operators to
arbitrary complex powers, gives a new stargenfunction with the expected
stargenvalues. The new stargenfunction in general should be a linear
combination of different branches, but we are not yet able to decompose it.
What makes the problem more complicated is that, when the plus and minus
branches degenerate, there exists an extra branch of stargenfunctions. Such
an extra branch is very likely to be activated by some combinations of
creation and annihilation operators. To illustrate this, in (\ref{aaftest})
we set $\lambda ^{+}=\lambda ^{-}=\tilde{\lambda}$, which leads to a
diagonal stargenfunction, and by performing the $\tau ^{\pm }$ integrals in (%
\ref{gllc}) we get 
\begin{eqnarray}
&&g\left( \lambda ,\tilde{\lambda},\tilde{\lambda},c\right)  \notag \\
&=&\int_{-1}^{1}ds\ 2^{-\tilde{\lambda}}\frac{\left( 1-s\right) ^{\lambda -%
\frac{1}{2}+\tilde{\lambda}}}{\left( 1+s\right) ^{\lambda +\frac{1}{2}-%
\tilde{\lambda}}}\left( \frac{c+s}{1-s^{2}}\right) ^{\tilde{\lambda}%
}e^{2sw}U\left( -\tilde{\lambda},1,\frac{2\left( 1-s^{2}\right) }{c+s}%
w\right) \text{ .}  \label{gspeculate}
\end{eqnarray}%
Note that the extra branch can be represented (except for the special cases
with half-integer stargenvalues) by the second term in (\ref{fsolnLU}),
which contains a factor $U\left( \frac{1}{2}-\lambda ,1,4w\right) $ that
logarithmically diverges as $w\rightarrow 0$. In (\ref{gspeculate}), we can
find a similar factor $U\left( -\tilde{\lambda},1,\frac{2\left(
1-s^{2}\right) }{c+s}w\right) $ whose expansion w.r.t.\ $w$ also gives $\log
w$, and we therefore speculate that the extra branch is involved.

\section{Conclusion and discussion}

\label{Sec conclusion}

In this paper, motivated by the construction of initial data in the
holomorphic gauge to solve Vasiliev's equations, we have studied the
stargenfunctions of the number operator, as defined in (\ref{originaleq})
with generic complex stargenvalues. We have re-written these
stargenfunctions using integral representations and also expressed the
creation and annihilation operators to complex powers by integrals via the
Mellin transform, so that their star-products can be computed conveniently.
We have in particular picked out a set of stargenfunctions (\ref{fsolnLL})
to investigate how they are changed by creation and annihilation operators.
This set of stargenfunctions can be written as linear combinations of two
subsets of stargenfunctions (the \textquotedblleft plus\textquotedblright\
and \textquotedblleft minus\textquotedblright\ branches), except that the
two subsets intersect when left and the right stargenvalues differ by an
integer. The whole plus (minus) branch of stargenfunctions can be related
(up to overall constants) by -- and is closed under -- the
star-multiplications with creation (annihilation) operators to generic
complex powers. However, by mixing the usage of creation and annihilation
operators, (\ref{fsolnLL}) does not seem to close, although the
stargenvalues are likely to be what we expect. We have confirmed that a
diagonal stargenfunction of the type (\ref{fsolnLL}), star-multiplied by two
factors respectively of creation and annihilation operators to complex
powers, yields another stargenfunction with the expected stargenvalues. We
speculate that the resulting stargenfunction should be a linear combination
of the plus and minus branches when they are non-degenerate and that another
extra branch should be involved in the degenerate case.

We are not yet able to use the above results to construct valid solutions to
Vasiliev's equations. One of the difficulties happens when we go beyond
linear combinations of the stargenfunctions. As in (\ref{ansatzholo}) the
twistor space gauge field $V^{\prime }$ is expressed in terms of
star-product series of the initial-data field $\Psi $. Thus, for example if
we use the stargenfunctions $f_{\lambda _{L}|\lambda _{R}}^{\pm }$ above to
construct $\Psi $, we will encounter star-multiplications to all orders
between these stargenfunctions, and we have to make sure that such
star-multiplications make sense. Let us take a look at the diagonal
stargenfunctions to illustrate the problem. When the stargenvalues are
half-integers, according to \cite{Iazeolla:2011cb}, $f_{\lambda |\lambda }$
should act like a projector: $f_{\lambda _{1}|\lambda _{1}}\star f_{\lambda
_{2}|\lambda _{2}}\propto \delta _{\lambda _{1}\lambda _{2}}f_{\lambda
_{2}|\lambda _{2}}$. Naively we expect the same thing to happen for generic
complex stargenvalues, which however, does not seem to be true.

If we multiply the expression (\ref{fllint}) by itself, by using the lemma 
\begin{equation}
e^{2sw}\star e^{2s^{\prime }w}=\frac{1}{1+ss^{\prime }}e^{2\frac{s+s^{\prime
}}{1+ss^{\prime }}w}\text{ ,}
\end{equation}%
and denoting $\Delta \lambda \equiv \lambda _{1}-\lambda _{2}$, we get 
\begin{eqnarray}
&&f_{\lambda _{1}|\lambda _{1}}\star f_{\lambda _{2}|\lambda _{2}}  \notag \\
&=&\int_{-1}^{1}\int_{-1}^{1}ds\ ds^{\prime }\ e^{2ws}\frac{\left(
1-s\right) ^{\lambda _{1}-\frac{1}{2}}}{\left( 1+s\right) ^{\lambda _{1}+%
\frac{1}{2}}}\frac{\left( 1-s^{\prime }\right) ^{\lambda _{2}-\frac{1}{2}}}{%
\left( 1+s^{\prime }\right) ^{\lambda _{2}+\frac{1}{2}}}\frac{1}{%
1+ss^{\prime }}e^{2\frac{s+s^{\prime }}{1+ss^{\prime }}w}  \notag \\
&=&\int_{-1}^{1}ds\frac{\left( 1-s\right) ^{\Delta \lambda -1}}{\left(
1+s\right) ^{\Delta \lambda +1}}\int_{-1}^{1}\ du\ \frac{\left( 1-u\right)
^{\lambda _{2}-\frac{1}{2}}}{\left( 1+u\right) ^{\lambda _{2}+\frac{1}{2}}}%
e^{2uw}  \notag \\
&=&f_{\lambda _{2}|\lambda _{2}}\int_{-1}^{1}ds\frac{\left( 1-s\right)
^{\Delta \lambda -1}}{\left( 1+s\right) ^{\Delta \lambda +1}}\text{ ,}
\label{fllsq}
\end{eqnarray}%
where in the second last step the integration variable $s^{\prime }$ has
been changed by $s^{\prime }=\frac{s-u}{su-1}$ or equivalently $u=\frac{%
s+s^{\prime }}{1+ss^{\prime }}$. In the last line of (\ref{fllsq}),
obviously the integral is problematic: let $\epsilon $ be a small positive
number, then we have%
\begin{equation}
\int_{-1+\epsilon }^{1-\epsilon }ds\frac{\left( 1-s\right) ^{\Delta \lambda
-1}}{\left( 1+s\right) ^{\Delta \lambda +1}}=\frac{1}{2\Delta \lambda }\left[
\left( \frac{2-\epsilon }{\epsilon }\right) ^{\Delta \lambda }-\left( \frac{%
\epsilon }{2-\epsilon }\right) ^{\Delta \lambda }\right] \text{ ,}
\label{inteps}
\end{equation}%
and as the factors $\epsilon ^{-\Delta \lambda }$ and $\epsilon ^{\Delta
\lambda }$ indicate, if $\epsilon $ is set to $0$, for any $\Delta \lambda $
the integral does not converge.\footnote{%
The case that $\Delta \lambda =0$ is a little bit special. By taking $\Delta
\lambda \rightarrow 0$ either before or after doing the integral, (\ref%
{inteps}) gives Ln$\left( \frac{2}{\epsilon }-1\right) $, which is divergent
for vanishing $\epsilon $.}

In the paper \cite{Iazeolla:2011cb} for half-integer stargenvalues, a
prescription has been given: the real-axis integral $\int_{-1}^{1}$ has been
replaced with the contour integral around either $s=1$ or $s=-1$ (both types
of integrals can produce the same Laguerre polynomial at the beginning). In
that case, while the real-axis integral is problematic, the contour one
converges and leads to the expected projector result. On the contrary, in
this paper for complex stargenvalues, we have to prescribe with a different
contour $C_{-1,1}$ as shown in Figure \ref{Cm1p1}, i.e.\ we now have the
factor:%
\begin{equation}
\oint_{C_{-1,1}}ds\frac{\left( s-1\right) ^{\Delta \lambda -1}}{\left(
s+1\right) ^{\Delta \lambda +1}}  \label{circlepres}
\end{equation}%
in the prescribed result of $f_{\lambda _{1}|\lambda _{1}}\star f_{\lambda
_{2}|\lambda _{2}}$. For $\Delta \lambda \neq 0$, we can do the indefinite
integral: 
\begin{equation}
\int ds\frac{\left( s-1\right) ^{\Delta \lambda -1}}{\left( s+1\right)
^{\Delta \lambda +1}}=\frac{1}{2\Delta \lambda }\left( \frac{s-1}{s+1}%
\right) ^{\Delta \lambda }+\ \text{const. .}  \label{indefint}
\end{equation}%
Within a given branch of the function on the r.h.s.\ of (\ref{indefint}),%
\footnote{%
The branch cut is the same as shown in Figure \ref{Cm1p1}.} by going through
a full cycle along the contour $C_{-1,1}$, both $\left( s-1\right) $ and $%
\left( s+1\right) $ acquire the same phase factors that cancel each other,
i.e.\ the value of the primitive function is not changed by a full cycle
along $C_{-1,1}$. Therefore, the integral (\ref{circlepres}) gives 0. For
the special case $\Delta \lambda =0$, (\ref{circlepres}) becomes the sum of
two cancelling residues around $s=1$ and $-1$, which also gives 0.
Therefore, with such a prescription, $f_{\lambda _{1}|\lambda _{1}}\star
f_{\lambda _{2}|\lambda _{2}}=0$ always holds, even if $\lambda _{1}=\lambda
_{2}$.\footnote{%
It is perhaps interesting to observe that setting $\Delta \lambda =0$ on the
l.h.s.\ of (\ref{indefint}) gives a finite primitive function $-$ArcCoth$%
\left( s\right) $, but the r.h.s.\ blows up when $\Delta \lambda \rightarrow
0$. We do not yet know whether this can be exploited in some way to
prescribe a non-zero result for (\ref{circlepres}).} It is not yet certain
to us how this should be interpreted -- perhaps we should for consistency
exclude half-integer stargenvalues when we use such stargenfunctions to
construct the initial data, or perhaps we should find a better prescription
that we are not yet aware of.\footnote{%
Note that here the stargenvalues have continuous spectra. A naive guess is
that, unlike the discrete case for half-integers, perhaps here with some
prescriptions the star-products between stargenfunctions should lead to
distributions like Dirac delta functions instead of Kronecker deltas, and
thus infinities and zeros could be somewhat expected. Furthermore, note that
in this paper we have not yet discussed much about the extra branch of
stargenfunctions in the case of degenerate plus and minus branches, or any
other special stargenfunctions corresponding to distributions in $w$. There
might be a chance that the plus and minus branches summed up together with
these additional branches could be normalized more easily.}

Another complication is the star-product with the inner Klein operators. As
introduced around (\ref{eigenfunc12}), to construct the initial data, we
need to make two copies of the stargenfunctions. If each copy has two
branches e.g.\ ($+,-$), then two copies give four branches ($++,+-,-+,--$)
in the initial data. A stargenfunction of any particular branch
star-multiplied by $\kappa _{y}$ or $\bar{\kappa}_{\bar{y}}$ in general may
lead to combinations of all branches. Thus the computation is very much
involved, and we will continue these investigations in our future work.

\bigskip

\bigskip

\noindent \textbf{Acknowledgement \ }\ This work is supported by
\textquotedblleft the Fundamental Research Funds for the Central
Universities, NO. NS2020054\textquotedblright\ of China. The author would
like to thank R. Aros, D. De Filippi, V.E. Didenko, C. Iazeolla, P. Sundell
and the anonymous referee for inspiring discussions. In particular, the
author is grateful to C. Iazeolla and P. Sundell for pointing out important
details about the contour integrals. The author would also like to thank all
the organizers and hosts of the APCTP-KHU Workshop on Higher Spin Gravity in
Pohang, South Korea, for hospitality and for providing this excellent
opportunity for discussions that helped the improvement of this paper.

%%%%%%%%%%%%%%%%%%%%%%%%%%%
\bibliographystyle{utphys}
\bibliography{biblio}
%%%%%%%%%%%%%%%%%%%%%%%%%%%

\end{document}